\documentclass{svmult}
\usepackage{makeidx}
\usepackage{graphicx}
\usepackage{multicol}

\makeindex

\begin{document}

\title*{\bf Hydrodynamics for a granular gas from an exactly solvable kinetic model}
\titlerunning{Hydrodynamics for a granular gas}
%for an abbreviated version of
\author{Aparna Baskaran and James W. Dufty}
% Use \authorrunning{Granular Gas Hydrodynamics} for an abbreviated version of
\institute{Department of Physics, University of Florida,
Gainesville, FL 32611, USA \texttt{aparna@phys.ufl.edu,
dufty@phys.ufl.edu}}
\maketitle

\begin{abstract}
A simple exactly solvable kinetic model for the non-linear
inelastic hard sphere Boltzmann equation is used to explore the
relevance of hydrodynamics for a granular gas. The equation
predicts a non-trivial homogeneous cooling state (HCS), including
algebraic decay at large velocities. The linearized kinetic
equation for small perturbations about the HCS is solved exactly.
It is shown that the hydrodynamic excitations exist in this
linearized dynamics, and their detailed form is shown to agree
with results from the Chapman-Enskog method up through
Navier-Stokes order. The existence of the hydrodynamic modes at
short wavelengths, far beyond Navier-Stokes order is demonstrated
as well. Finally, the precise sense in which the hydrodynamic
excitations dominate the dynamics at long times is
described.\footnote{ This research was presented at the
''Modelling and Numerics of Kinetic Dissipative Systems''
workshop, Lipari, Italy, June 2004. An abbreviated version of the
present work has been submitted for publication in the conference
proceedings (L. Pareschi, G. Russo, G. Toscani Eds., Nova Science,
New York.)}
\end{abstract}

%for an abbreviated version of

% Use \authorrunning{Granular Gas Hydrodynamics} for an abbreviated version of

\section{Introduction}

The applicability of a hydrodynamic description for a granular gas has been
an issue of considerable interest in recent years \cite
{IGoldgg,BDMj2003,DBlip}. The concepts involved and the precise questions to
be asked have been clarified to a large extent in recent works \cite
{DBlip,BDmodes03}, and partial answers to these questions have been given.
The setting for all of these discussions has been the inelastic hard sphere
Boltzmann equation. However, direct demonstration of the transition to a
hydrodynamic description is made difficult by the analytic complexity of the
Boltzmann collision operator. The objective here is to provide such an
illustration by addressing these issues in the more limited context of a
simple and exactly solvable model kinetic equation for a granular gas \cite
{BMD96,DBZ03}. This model has the primary characteristics of the Boltzmann
kinetic equation: dissipation of energy, a non-trivial homogeneous cooling
state (HCS), and exact macroscopic balance equations for the hydrodynamic
fields. Hence it is a good starting point for a demonstration of principle
in the context of granular gases.

A closed set of equations for the density, flow velocity and the granular
temperature can be obtained from both the Boltzmann kinetic theory and the
model kinetic theory considered here by the Chapman-Enskog procedure \cite
{BDKS97}. The drawback of this procedure is its practical limitation to
states with small spatial gradients, and the assumption of a special form
for the solution (a ''normal'' solution \cite{DBlip}). There remain
questions about the space and time scales on which such a solution can be
expected. Also, the lack of energy conservation has led some to question
whether the local energy or temperature should properly be considered a
''slow mode'' of the system that can dominate at long times.

A direct probe of these questions is possible by characterizing all the
excitations in the linearized dynamics of the system for small spatial
perturbations of a homogeneous state, and looking for a time scale
separation between the hydrodynamic excitations and the other microscopic
excitations \cite{BDmodes03}. This prescription is difficult to carry out
for the hard-sphere Boltzmann equation because of the complicated collision
operator involved, even at the linearized level. However, this linear
problem can be solved exactly for the model under consideration here. It is
found that 1) hydrodynamic excitations exist and agree with those obtained
from the Chapman-Enskog method under the same conditions of long
wavelengths, 2) the hydrodynamic excitations exist over a range of shorter
wavelengths far beyond the restrictions of the Navier-Stokes approximation,
and 3) the other dynamics (called ''microscopic'' excitations) typically
decay in a few collision times so that the hydrodynamic excitations dominate
at long times. Some qualifications of 3) are noted. These conclusions hold
even for conditions of strong dissipation.

The paper is organized as follows. First, the definition of the
kinetic model is recalled and its solution for spatially
homogeneous states is given. It is noted that a special HCS
solution results for a wide class of initial conditions after a
few collisions times. Next, weakly inhomogeneous states for small
perturbations about this HCS are considered. First, the
hydrodynamic description obtained from the Chapman-Enskog results
is recalled in order to define "hydrodynamic" excitations. Next,
the kinetic equation is linearized and solved exactly by
Fourier-Laplace transformation. The analytic properties in the
complex plane determine all possible excitations, and it is shown
that the hydrodynamic excitations are among these. Finally, the
general solution is studied in the long wavelength limit where
hydrodynamic and microscopic excitations can be displayed more
explicitly. Conditions for the dominance of the hydrodynamic
excitations at long times is considered in some detail. The
results and remaining questions are summarized in the last
section.

\section{Definition of the kinetic model}

A granular gas at low density is found to be well described by a
system of $ N $ smooth hard spheres undergoing inelastic
collisions parameterized by a coefficient of restitution $\alpha
$. The number density for particles with position $\mathbf{r}$ and
velocity $\mathbf{v}$ at time $t$ is determined from the Boltzmann
equation \cite{vNyE01,DuftACS01}
\begin{equation}
\left( \frac{\partial }{\partial t}+\mathbf{v}\cdot \nabla \right) f(\mathbf{
r},\mathbf{v},t)=J\left( \mathbf{r},\mathbf{v}|f(t)\right) .  \label{2.1}
\end{equation}
The detailed form of the collision operator $J\left( \mathbf{r},\mathbf{v}
|f(t)\right) $ is not required for the purposes here, only some of its
properties. Among these are its moments with respect to $1,$ $\mathbf{v}$,
and $v^{2}$
\begin{equation}
\int d\mathbf{v}\left(
\begin{array}{c}
1 \\
\mathbf{v} \\
\frac{1}{2}m\left( \mathbf{v}-\mathbf{u}\right) ^{2}
\end{array}
\right) J\left( \mathbf{r},\mathbf{v}|f(t)\right) =\left(
\begin{array}{c}
0 \\
\mathbf{0} \\
-\frac{3}{2}nT\zeta
\end{array}
\right) .  \label{2.2}
\end{equation}
Here $n$, $\mathbf{u}$, and $T$ are the local density, flow velocity, and
temperature at time $t$, obtained from the distribution function according
to
\begin{equation}
\left(
\begin{array}{c}
n(\mathbf{r},t) \\
n(\mathbf{r},t)\mathbf{u}(\mathbf{r},t) \\
\frac{3}{2}n(\mathbf{r},t)T(\mathbf{r},t)
\end{array}
\right) =\int d\mathbf{v}\left(
\begin{array}{c}
1 \\
\mathbf{v} \\
\frac{1}{2}m\left( \mathbf{v}-\mathbf{u}\right) ^{2}
\end{array}
\right) f(\mathbf{r},\mathbf{v,}t).
\end{equation}
Finally, $\zeta $ is the ''cooling rate'' (see Eq.(\ref{3.2})
below to justify the terminology)
\begin{equation}
\zeta \left( \mathbf{r},t\right) =\left( 1-\alpha ^{2}\right) \frac{m\pi
\sigma ^{2}}{24n\left( \mathbf{r},t\right) T\left( \mathbf{r},t\right) }\int
\,d\mathbf{v}\,\int \,d\mathbf{v}_{1}\,g^{3}\ f(\mathbf{r},\mathbf{v},t,)f(
\mathbf{r},\mathbf{v}_{1},t).  \label{2.4}
\end{equation}
The two zeros in Eq.(\ref{2.2}) correspond to the conservation of mass and
momentum respectively. The source in the third term implies that the energy
is not conserved and the system is ''cooling'' (note that it vanishes in the
elastic limit of $\alpha \rightarrow 1$). $\ $The functions $n$, $\mathbf{u}$
, and $T$ are the hydrodynamic fields, and the properties (\ref{2.2})
together with the Boltzmann equation give the exact macroscopic balance
equations relating these fields. These balance equations are the essential
starting point for the formulation of a hydrodynamic description for the
system. Consequently, the properties (\ref{2.2}) should be preserved by any
model kinetic theory intended to explore hydrodynamics.

The kinetic model considered here is a simple extension of the BGK
model for gases with elastic collisions \cite{BGK54}, to
capture the properties of a granular gas. It is obtained by
replacing the bilinear collision operator in the Boltzmann
equation by \cite{BMD96}
\begin{equation}
J\left( \mathbf{r},\mathbf{v}|f(t)\right) \rightarrow -\nu \left( f-g\right)
,  \label{2.5}
\end{equation}
where $\nu $ is a velocity independent parameter of the model and $g$ is
chosen to be a functional of $f$ such that the moment conditions (\ref{2.2})
are satisfied. As in the BGK model $g$ is taken to be a Gaussian
\begin{equation}
g(\mathbf{r},\mathbf{V},t\mid f)=A(\mathbf{r},t)e^{-B(\mathbf{r},t)V^{2}},
\end{equation}
where $\mathbf{V}=\mathbf{v}-\mathbf{u(r,}t\mathbf{)}$ is the peculiar
velocity. $A$ and $B$ are chosen so as to enforce the moment conditions
above. They are found to be
\begin{equation}
A(\mathbf{r},t)=n(\mathbf{r},t)\left( \frac{B(\mathbf{r},t)}{\pi }\right)
^{3/2},\ \ B(\mathbf{r},t)=\frac{m}{2T(\mathbf{r},t)\left( 1-\frac{\zeta (
\mathbf{r},t)}{\nu (\mathbf{r},t)}\right) },  \label{2.7}
\end{equation}
where $m$ is the mass. In the elastic limit $g$ becomes the local
Maxwellian; otherwise, the effective temperature is changed to account for
the cooling implied by Eq. (\ref{2.2}).

It remains to choose the functions $\nu (\mathbf{r},t)$ and $\zeta (\mathbf{r
},t)$. In principle, these are specific functionals of $f$ in the Boltzmann
equation. Here, they are taken to depend on $f$ only through the temperature
and density. The cooling rate is chosen to be the same as that obtained from
the Boltzmann equation by using a local Maxwellian for $f$ in Eq.(\ref{2.4})
\cite{BMD96}. The collision frequency $\nu $ is chosen so as to fit one of
the Chapman-Enskog transport coefficients of the model to that of the
corresponding result from the Boltzmann equation, in this case the shear
viscosity $\eta $ (for details see \cite{DBZ03}). These choices give

\begin{equation}
\zeta (\mathbf{r},t)=\frac{5}{12}\left( 1-\alpha ^{2}\right) \nu _{0}(
\mathbf{r},t),\hspace{0.3in}\nu (\mathbf{r},t)=\left( 1-\frac{1}{4}(1-\alpha
)^{2}\right) \nu _{0}(\mathbf{r},t),
\end{equation}
where $\nu _{0}(\mathbf{r},t)$ is an average local collision frequency
\begin{equation}
\nu _{0}(\mathbf{r},t)=\frac{16}{5}n(\mathbf{r},t)\sigma ^{2}\sqrt{\frac{\pi
T(\mathbf{r},t)}{m}}.  \label{2.9}
\end{equation}
This completes the definition of the model kinetic equation
\begin{equation}
\left( \frac{\partial }{\partial t}+\mathbf{v}\cdot \nabla \right) f=-\nu
\left( f-g\right) .  \label{2.10}
\end{equation}
This is deceptively simple, since the right side is a highly nonlinear
functional of $f$ through its dependence on the hydrodynamics fields in $\nu
$ and $g$..

\section{Spatially homogeneous states}

As in the case of the Boltzmann equation, there exists no spatially
homogeneous steady solution to (\ref{2.10}) for the isolated system. This
can be seen by taking the second velocity moment of the kinetic equation to
get
\begin{equation}
T^{-1}\partial _{t}T=-\zeta .  \label{3.2}
\end{equation}
This implies that one of the moments of the distribution function is always
time dependent and hence there is no steady solution. Instead, there is a
special scaling solution, the HCS, which is approached in a few collision
times by most spatially homogeneous initial conditions.

Consider a general homogeneous initial distribution and look for solutions
to the model kinetic equation in the dimensionless form
\begin{equation}
f\left( \mathbf{v},t\right) =nv_{hcs}^{-3}f^{\ast }\left( \mathbf{v}^{\ast
},s\right) ,\hspace{0.4cm}\mathbf{v}^{\ast }=\mathbf{v}/v_{hcs}\left(
t\right)  \label{3.3}
\end{equation}
Here the velocity scaling is generated by the time dependent
thermal velocity and $s$ is an average collision number
\begin{equation}
v_{hcs}(t)=\sqrt{2T(t)/m},\hspace{0.3in}ds=\nu _{0}(t)dt.  \label{3.4}
\end{equation}
The dimensionless form for the model kinetic equation becomes
\begin{equation}
\left( \partial _{s}+\frac{1}{2}\zeta ^{\ast }\nabla _{\mathbf{v}^{\ast
}}\cdot \mathbf{v}^{\ast }+\nu ^{\ast }\right) f^{\ast }\left( \mathbf{v}
^{\ast },s\right) =\nu ^{\ast }g^{\ast }\left( \mathbf{v}^{\ast }\mid
f^{\ast }\right) .  \label{3.6}
\end{equation}
This can be integrated directly for initial homogeneous states to get the
general solution (see Appendix A and \cite{DBZ03})
\begin{eqnarray}
f^{\ast }(\mathbf{v}^{\ast },s) &=&e^{-\left( \frac{3}{2}\zeta ^{\ast }+\nu
^{\ast }\right) s}f^{\ast }(e^{-\frac{1}{2}\zeta ^{\ast }s}\mathbf{v}^{\ast
},0)+\nu ^{\ast }\left( \pi \left( 1-\frac{\zeta ^{\ast }}{\nu ^{\ast }}
\right) \right) ^{-3/2}  \nonumber \\
&&\times \int_{0}^{s}ds^{\prime }e^{-\left( \frac{3}{2}\zeta ^{\ast }+\nu
^{\ast }\right) s^{\prime }}\exp \left( -\frac{e^{-\zeta ^{\ast }s^{\prime
}}v^{\ast 2}}{1-\frac{\zeta ^{\ast }}{\nu ^{\ast }}}\right) ,  \label{3.9}
\end{eqnarray}
with the dimensionless constants $\nu ^{\ast }=\nu /\nu _{0},$ and
$\zeta ^{\ast }=\zeta /\nu _{0}$. Since the state is homogeneous
the flow velocity has been taken to vanish, $\mathbf{u}=0$, by
making the appropriate Galilean transformation.

The constant $\nu ^{\ast }$ is of order unity, so the domain of
integration in\ (\ref{3.9}) is exponentially bounded for $s>1$ and
for large $s$ the integral becomes independent of $s$. The first
term vanishes exponentially fast for initial conditions uniformly
bounded in the sense $f^{\ast }(e^{- \frac{1}{2}\zeta ^{\ast
}s}\mathbf{v}^{\ast },0)<C$. Therefore an $s -$ independent HCS
solution is obtained
\begin{equation}
f_{hcs}^{\ast} (v^{\ast })=\nu ^{\ast }\left( \pi \left( 1-\frac{\zeta
^{\ast }}{\nu ^{\ast }}\right) \right) ^{-3/2}\int_{0}^{\infty }ds^{\prime
}e^{-\left( \frac{3}{2}\zeta ^{\ast }+\nu ^{\ast }\right) s^{\prime }}\exp
\left( -\frac{e^{-\zeta ^{\ast }s^{\prime }}v^{\ast 2}}{1-\frac{\zeta ^{\ast
}}{\nu ^{\ast }}}\right)  \label{3.10}
\end{equation}
In terms of the dimensionless variables chosen, the HCS behaves as a
stationary state. It is universal in the same sense as the Maxwellian for
elastic collisions, since most initial homogeneous states evolve to it after
a few collisions. In the following it will be chosen as the reference state
to study small spatial perturbations of a homogeneous state. Clearly, the
results obtained will be similar for other homogeneous reference states, but
with an additional transient period as the reference state itself approaches
the HCS.

\section{Hydrodynamic excitations}

Hydrodynamic excitations are most easily defined in the long wavelength
limit where the Navier-Stokes hydrodynamic equations apply. Consider
inhomogeneous states characterized by smooth spatial and temporal variations
in the density, temperature and flow velocity, and for which the deviations
of the hydrodynamic fields from their values in the HCS are small (linear
hydrodynamics). The detailed form of these equations and the transport
coefficients occurring in them have been obtained from the kinetic equation
(Boltzmann and kinetic model) by the Chapman-Enskog method \cite
{BMD96,BDKS97} and are found to be
\begin{equation}
\partial _{t}\delta n+n_{hcs}\nabla \cdot \delta \mathbf{u}=0,  \label{4.1}
\end{equation}
\begin{eqnarray}
\partial _{t}\delta u_{i}+\frac{1}{n_{hcs}m}\left( n_{hcs}\partial
_{i}\delta T+T_{hcs}\partial _{i}\delta n\right)  \nonumber \\
-\frac{1}{n_{hcs}m}\left( \eta \left( \partial _{j}\delta
u_{i}+\partial _{i}\delta u_{j}-\frac{2}{3} \delta _{ij}\nabla
\cdot \delta \mathbf{u}\right) \right) =0,  \label{4.2}
\end{eqnarray}
\begin{eqnarray}
\partial _{t}\delta T+\frac{2}{3}T_{hcs}\nabla \cdot \delta \mathbf{u}
+\frac{3}{2}\zeta _{hcs}\delta T+\frac{T_{hcs}\zeta
_{hcs}}{n_{hcs}}\delta n \nonumber \\
-\frac{ 2}{3n_{hcs}}\left(
\kappa \nabla ^{2}\delta T+\mu \nabla ^{2}\delta n\right)=0,
\label{4.3}
\end{eqnarray}
where $\delta x=x-x_{hcs}$ denotes the deviation of a hydrodynamic
field from its value in the HCS. The transport coefficients $\eta
, \kappa, $and $\mu $ occurring in the above equations have been
evaluated in reference \cite{BMD96} .

The linear Navier-Stokes equations are complicated only by time
dependent coefficients due to the cooling reference state. This
time dependence can be eliminated by introducing the dimensionless
variables, $\delta n^{\ast }=\delta n/n_{hcs}$, $\delta T^{\ast
}=\delta T/T_{hcs},$ and $\delta \mathbf{ u}^{\ast }=\delta
\mathbf{u}/v_{hcs}$. In addition the time is transformed to the
collision number $s$ in (\ref{3.4}) and the dimensionless space
variable is $\mathbf{r}^{\ast }=\mathbf{r}\nu _{0}/v_{hcs}$.
Finally, since the equations are linear, it is sufficient to
consider a single Fourier component
\[
y_{\alpha }^{\ast }(\mathbf{r},s)=e^{i\mathbf{k\cdot r}^{\ast }}\tilde{y}
_{\alpha }^{\ast }(\mathbf{k},s)
\]
where the hydrodynamic fields are chosen to be
\begin{equation}
\tilde{y}_{\alpha }^{\ast }(\mathbf{k},s)\leftrightarrow \left( \delta
n^{\ast }(\mathbf{k},s),\delta T^{\ast }(\mathbf{k},s),\delta \mathbf{u}
^{\ast }(\mathbf{k},s)\cdot \widehat{\mathbf{k}},\delta \mathbf{u}^{\ast }(
\mathbf{k},s)\cdot \widehat{\mathbf{e}}_{1},\delta \mathbf{u}^{\ast }(
\mathbf{k},s)\cdot \widehat{\mathbf{e}}_{2}\right) .  \label{4.4}
\end{equation}
The components of the flow field are taken to be those along the orthonormal
triplet $\{\widehat{\mathbf{k}},\widehat{\mathbf{e}}_{1},\widehat{\mathbf{e}}
_{2}\}$. The dimensionless linear Navier-Stokes equations then become
\begin{equation}
\partial _{s}\tilde{y}_{\alpha }^{\ast }(\mathbf{k},s)+\mathcal{K}_{\alpha
\beta }\tilde{y}_{\alpha }^{\ast }(\mathbf{k},s)=0,  \label{4.6}
\end{equation}
with
\begin{equation}
 \mathcal{K}=\left(
\begin{array}{ccccc}
0 & 0 & ik & 0 & 0 \\
\zeta ^{\ast }+\frac{5}{4}\mu ^{\ast }k^{2} & \frac{\zeta ^{\ast }}{2}+\frac{
5}{4}\kappa ^{\ast }k^{2} & \frac{2}{3}ik & 0 & 0 \\
i\frac{k}{2} & i\frac{k}{2} & -\frac{\zeta ^{\ast }}{2}+\frac{2}{3}\eta
^{\ast }k^{2} & 0 & 0 \\
0 & 0 & 0 & -\frac{\zeta ^{\ast }}{2}+\frac{1}{2}\eta ^{\ast }k^{2} & 0 \\
0 & 0 & 0 & 0 & -\frac{\zeta ^{\ast }}{2}+\frac{1}{2}\eta ^{\ast }k^{2}
\end{array}
\right) .  \label{4.7}
\end{equation}
The transport coefficients are scaled with respect to their values
in the elastic limit, $\kappa ^{\ast }=\kappa /\kappa _{0},$ $\eta
^{\ast }=\eta /\eta _{0},$ $\mu ^{\ast }=\mu n_{hcs}T_{hcs}/\kappa
_{0}$, where the elastic limit values are $\kappa _{0}=15\eta
_{0}/4m$ and $\eta _{0}=5(mT_{hcs})^{1/2}/16\sigma ^{2}\pi
^{1/2}$. The dependence of $\eta ,$ $ \kappa, $ and $\mu $ on the
restitution coefficient $\alpha $ has been evaluated in reference
\cite{BMD96}.

The initial value problem can be solved directly with the solution
\begin{equation}
\tilde{y}^{*}_{\alpha }(\mathbf{k},s)=\sum_{\beta }C_{\alpha \beta
}e^{z_{\beta }(k)s}.  \label{4.12}
\end{equation}
Here $C_{\alpha \beta }$ are constants and $z_{\beta }(k)$ are the
eigenvalues of $\mathcal{K}\left( k\right) $ determined from
\begin{equation}
\det \left( z(k)\mathbf{1+}\mathcal{K}(k)\right) =0.  \label{4.9}
\end{equation}
As the Navier-Stokes equations are valid only up through order
$k^{2}$ the solutions to this equation are relevant only to the
same order
\begin{eqnarray}
z_{\alpha }(k) &\rightarrow &\left\{ -\frac{1}{\zeta ^{\ast }}k^{2},-\frac{
\zeta ^{\ast }}{2}+\left( \frac{2}{3\zeta ^{\ast }}-\frac{5}{4}\kappa ^{\ast
}\right) k^{2},\frac{\zeta ^{\ast }}{2}-\left( \frac{1}{3\zeta ^{\ast }}+
\frac{2}{3}\eta ^{\ast }\right) k^{2},\right.  \nonumber \\
&&\left. \frac{\zeta ^{\ast }}{2}-\frac{1}{2}\eta ^{\ast }k^{2},\frac{\zeta
^{\ast }}{2}-\frac{1}{2}\eta ^{\ast }k^{2}\right\} .  \label{4.10}
\end{eqnarray}
For the discussion below it is noted that in particular, at $k=0$, these
become
\begin{equation}
z_{\alpha }(0)\rightarrow \left\{ 0,-\frac{\zeta ^{\ast }}{2},\frac{\zeta
^{\ast }}{2},\frac{\zeta ^{\ast }}{2},\frac{\zeta ^{\ast }}{2}\right\} .
\label{4.11}
\end{equation}

The hydrodynamic excitations can now be defined precisely. They
are the excitations of the form (\ref{4.12}) with $z_{\beta }(k)$
defined to be the frequencies that are continuously connected to
$z_{\alpha }(0)$ defined in (\ref{4.11}) as $k$ goes to zero.
When such modes can be identified it is said that hydrodynamic
excitations exist. In that case, an internal consistency check is
the agreement of those modes with the Chapman-Enskog results (\ref
{4.10}) at order $k^{2}$. The modes $z_{4}(k)=z_{5}(k)$ are
associated with the components of the flow field orthogonal to the
$\mathbf{k }$. These are called transverse shear modes. The other
three modes are called longitudinal modes. Since $\mathcal{K}$ is
block diagonal the transverse modes decouple from the other modes
and hence can be treated more easily and directly.

\section{Linear kinetic theory}

In the rest of the paper, the dimensionless variables introduced
in the previous sections are used throughout and the asterisk is
deleted for simplicity of notation. To describe small
perturbations of the distribution function about the HCS solution,
define $\delta f$ by
\begin{equation}
f=f_{hcs}+\delta f.  \label{5.1}
\end{equation}
Substitution of this form into the kinetic equation (\ref{2.10})
and retaining only linear order terms in $\delta f$ defines the
linear kinetic theory. The model collision operator of
(\ref{2.10}) depends on the distribution function explicitly and
implicitly through the hydrodynamic fields in $\nu $ and $g$. The
linearized kinetic equation is identified in Appendix B and is
found to be
\begin{equation}
\left( \partial _{s}+\mathbf{v}\cdot \nabla _{\mathbf{r}}+\nu +\frac{1}{2}
\zeta \nabla _{\mathbf{v}}\cdot \mathbf{v}\right) \delta f=M_{\alpha
}y_{\alpha }  \label{5.2}
\end{equation}
where
\begin{equation}
M_{a}(\mathbf{v})=\left(
\begin{array}{c}
\left( \zeta \nabla _{\mathbf{v}}\cdot \mathbf{v}+\nu \right) f_{hcs} \\
\frac{1}{2}\nabla _{\mathbf{v}}\cdot \mathbf{v}\left( \frac{1}{2}\zeta
-\left(\frac{1}{2} \zeta \nabla _{\mathbf{v}}\cdot \mathbf{v}+\nu \right) \right)
f_{hcs} \\
\left( -\frac{1}{2}\zeta -\left(\frac{1}{2} \zeta \nabla _{\mathbf{v}}\cdot \mathbf{v}
+\nu \right) \right) \nabla _{\mathbf{v}}f_{hcs}
\end{array}
\right) .  \label{5.3}
\end{equation}
The hydrodynamic fields $y_{\alpha }$ are defined in (\ref{4.4}).
The kinetic equation can be solved by a Fourier-Laplace
transformation
\begin{equation}
\delta \tilde{\tilde{f}}(\mathbf{k},\mathbf{v},z)=\int_{0}^{\infty
}dse^{-zs}\int d\mathbf{r}e^{i\mathbf{k\cdot r}}\delta f(\mathbf{r},\mathbf{v
},s).  \label{5.4}
\end{equation}
The formal solution is
\begin{equation}
\delta \tilde{\tilde{f}}(\mathbf{k},\mathbf{v},z)=\mathcal{R}\delta \tilde{f}
(\mathbf{k},\mathbf{v},0)+\tilde{\tilde{y}}_{\alpha }(\mathbf{k},z)\mathcal{R
}M_{\alpha }(\mathbf{v})  \label{5.5}
\end{equation}
where the resolvent operator $\mathcal{R}$ is
\begin{equation}
\mathcal{R}(\mathbf{k},z)=\left( z+\nu +i\mathbf{k}\cdot \mathbf{v}+\frac{1}{
2}\zeta \nabla _{\mathbf{v}}\cdot \mathbf{v}\right) ^{-1}  \label{5.6}
\end{equation}

This is an implicit solution for $\delta
\tilde{f}(\mathbf{k},\mathbf{v},z)$ because the hydrodynamic
fields $\left\{ \tilde{\tilde{y}}_{\alpha }\right\} $ are linear
functionals of $\delta \tilde{\tilde{f}}$
\begin{equation}
\tilde{\tilde{y}}_{\alpha }(\mathbf{k},z)=\int d\mathbf{v}\phi _{\alpha
}\left( \mathbf{v}\right) \delta \tilde{\tilde{f}}(\mathbf{k},\mathbf{v},z).
\label{5.7}
\end{equation}
where
\begin{equation}
\phi _{\alpha }=\left( 1,\left( \frac{2}{3}v^{2}-1\right) ,\mathbf{v}\cdot
\widehat{\mathbf{k}},\mathbf{v}\cdot \widehat{\mathbf{e}}_{1},\mathbf{v}
\cdot \widehat{\mathbf{e}}_{2}\right)  \label{5.8}
\end{equation}
The $\left\{ \tilde{\tilde{y}}_{\alpha }\right\} $ can be
determined self-consistently using (\ref{5.5}) in (\ref{5.7}) to
get
\begin{equation}
\tilde{\tilde{y}}_{\alpha }(\mathbf{k},z)=(\mathbf{1}-Q(\mathbf{k}
,z))_{\alpha \beta }^{-1}I_{\beta }(\mathbf{k},z)
\label{5.9}
\end{equation}
where
\begin{equation}
I_{\alpha }=\int d\mathbf{v}\phi _{\alpha }( \mathbf{v})
\mathcal{R}\delta \tilde{f}(\mathbf{k},\mathbf{v},0),\hspace{0.4cm }
Q_{\alpha \beta }=\int d\mathbf{v}\phi _{\alpha }(\mathbf{v})\mathcal{R}
M_{\beta }.  \label{5.10}
\end{equation}
Finally, substitution of (\ref{5.9}) in (\ref{5.5}) gives the desired
solution
\begin{equation}
\delta \tilde{\tilde{f}}(\mathbf{k},\mathbf{v},z)=\mathcal{R}\delta \tilde{f}
(\mathbf{k},\mathbf{v},0)+\mathcal{R}M_{\alpha }(\mathbf{v})(\mathbf{1}
-Q)_{\alpha \beta }^{-1}I_{\beta }(\mathbf{k},z).
\label{5.11}
\end{equation}

It remains to make explicit the action of the resolvent operator
in (\ref {5.11}). This is determined in Appendix A with the
result, for an arbitrary function of the velocity $X\left(
\mathbf{v}\right)$,
\begin{equation}
\mathcal{R}X\left( \mathbf{v}\right) =\int_{0}^{\infty }dse^{-\left( z+\nu +
\frac{3}{2}\zeta \right) s}\exp \left( \frac{2}{\zeta }\left( e^{-\frac{1}{2}
\zeta s}-1\right) i\mathbf{k}\cdot \mathbf{v}\right) X\left( e^{-\frac{1}{2}
\zeta s}\mathbf{v}\right) ,  \label{5.12}
\end{equation}
In this way the general linear solution (\ref{5.11}) is reduced to
quadratures. $I$ and $Q$ in Eq.(\ref{5.10}) are evaluated in
Appendix C.

\subsection{Transverse modes}

The analysis for transverse shear mode excitations is simplest and
illustrates the more general case. Consider an initial perturbation in the
transverse flow velocity only
\begin{equation}
\delta \tilde{f}(\mathbf{k},\mathbf{v},0)=-\widehat{\mathbf{e}}_{1}\cdot
\nabla _{\mathbf{v}}f_{hcs}\left( v\right) \delta \tilde{u}_{t}(\mathbf{k}
,0) \hspace{0.3in}u_{t}=\mathbf{u\cdot }\widehat{\mathbf{e}}_{1}  \label{6.1}
\end{equation}
This perturbation results from a small change in the velocity of
the HCS along the transverse direction $\widehat{\mathbf{e}}_{1}$.
It couples only to $\mathbf{u\cdot }\widehat{\mathbf{e}}_{1}$ (due
to symmetry) and Eq. (\ref{5.9}) for the response of that field
reduces to
\begin{equation}
\delta \tilde{\tilde{u}}_{t}(\mathbf{k},z)=\frac{I(\mathbf{k}
,z)}{1-Q_{44}(\mathbf{k},z)}\delta \tilde{u}_{t}(\mathbf{k},0).  \label{6.2}
\end{equation}
with
$I(\mathbf{k},z)=I_{4}(\mathbf{k}
,z)/\delta \tilde{u}_{t}(\mathbf{k},0)$. The functions
$I( \mathbf{k},z)$and $Q_{44}(\mathbf{k},z)$ are
evaluated using the property ( \ref{5.12}) with the results
\begin{equation}
I(k,z)=\int_{0}^{\infty }dse^{-\left( z+\nu-\frac{1}{2}\zeta
\right) s}\nu \int_{0}^{\infty }ds^{\prime }e^{-\nu s^{\prime }}\allowbreak
\exp \left( -e^{\zeta s^{\prime }}a(s)k^{2}\right) ,  \label{6.3}
\end{equation}
where
\begin{equation}
a(s)=\left( 1-\frac{\zeta }{\nu }\right) \left( \frac{e^{\frac{1}{2}\zeta
s}-1}{\zeta }\right) ^{2}.  \label{6.3a}
\end{equation}
Similarly, $Q_{44}(k,z)$is found to be
\begin{equation}
Q_{44}(k,z)=\left( \nu -\zeta k^{2}\frac{d}{dk^{2}}\right) \tilde{\tilde{I}}
(k,z)  \label{6.4}
\end{equation}
Consider first the long wavelength limit $k\rightarrow 0$
\begin{equation}
I(\mathbf{0},z)=\frac{1}{z+\nu -\frac{\zeta }{2}},\hspace{
0.3in}1-Q(\mathbf{0},z)=\frac{z-\frac{\zeta }{2}}{z+\nu -\frac{\zeta }{2}}
\hspace{0.3in}  \label{6.6}
\end{equation}
\begin{equation}
\delta \tilde{\tilde{u}}_{t}(\mathbf{k},z)\rightarrow \frac{1}{z-\frac{1}{2}
\zeta }\delta \tilde{u}_{t}(\mathbf{k},0).  \label{6.7}
\end{equation}
Comparison with (\ref{4.11}) shows that this is the expected hydrodynamic
pole, confirming the existence of hydrodynamics in the kinetic theory for
response to a transverse perturbation.

The hydrodynamic pole in the long wavelength limit arises from the
zero of $ 1-Q_{44}(\mathbf{0},z(0)).$ Its continuation to $z(k)$
at finite $k$ is therefore defined to be the solution to
\begin{equation}
1-Q_{44}(\mathbf{k},z(k))=0,\hspace{0.3in}  \label{6.8}
\end{equation}
with the condition $z(0)=\zeta /2$ (the poles of the function
$\tilde{\tilde{I }}(\mathbf{k},z)$ determine the microscopic
excitations in $\delta \tilde{ \tilde{u}}_{t}(\mathbf{k},z)$ for
finite $k$). It is easily shown that $z(k)$ is a real function of
$k^{2}$. The integrals above are defined only for $z> \frac{\zeta
}{2}-\nu.$ Therefore, the hydrodynamic pole can exist only for
values of $k$ such that $z\left( k\right) >-\left( \nu -\frac{
\zeta }{2} \right).$

\begin{figure}[tbp]
\centering
%Use the relevant command for your figure-insertion program
\includegraphics[height=8cm]{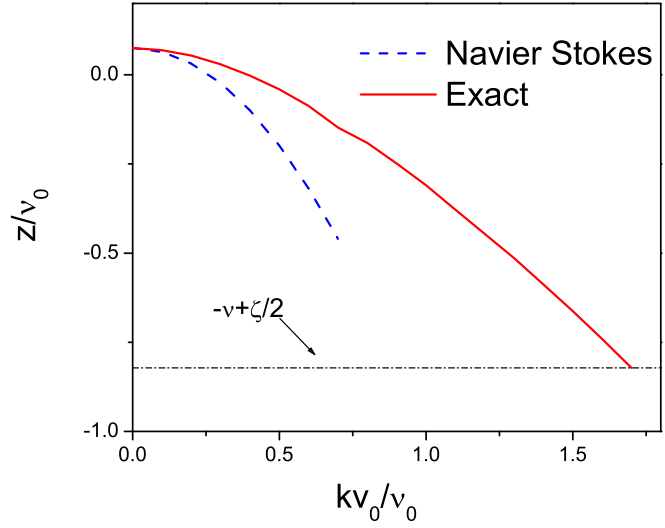} %
%\picplace{6cm}{4cm} % Give the correct figure height and width in cm
%
% Give a unique label
\caption{Transverse hydrodynamic mode}
\label{fig:1}
\end{figure}
Figure 1 illustrates the numerical solution to (\ref{6.8}) for $\alpha =0.8$%
, showing the hydrodynamic pole exists for all $k<k_{m}$, where
$z\left( k_{m}\right) =\frac{\zeta}{2}-\nu$. For small $k$ direct
expansion of (\ref {6.8}) to order $k^{2}$ gives the solution
\begin{equation}
z(k)\rightarrow \frac{1}{2}\zeta (\alpha )-\frac{1}{2}\eta (\alpha )k^{2}.
\label{6.9}
\end{equation}
This is the same form as that for the transverse hydrodynamic mode obtained
from the Navier-Stokes equation in (\ref{4.10}). The shear viscosity
obtained here is
\begin{equation}
\eta (\alpha )=\frac{1}{\left( \nu(\alpha )-\frac{1}{2}\zeta (\alpha
)\right) },  \label{6.10}
\end{equation}
which is the same as that found by the Chapman-Enskog method in \cite{BMD96}
and hence is an independent verification of that method. Interestingly,
Figure 1 suggests that the Navier-Stokes approximation begins to fail for
relatively small wavevectors.

\subsection{Longitudinal modes}

Return now to the general solution (\ref{5.5}) and consider
initial conditions that generate the perturbations $\delta
\tilde{n}(\mathbf{k},0),$ $ \delta \tilde{T}(\mathbf{k},0),$ and
$\delta \tilde{\mathbf{u}}(\mathbf{k} ,0)\cdot
\widehat{\mathbf{k}}$. The Laplace transform of these fields at
later times is given by (\ref{5.9})
\begin{equation}
\tilde{\tilde{y}}_{\alpha }(\mathbf{k},z)=(\mathbf{1}-Q(\mathbf{k}
,z))_{\alpha \beta }^{-1}I_{\beta }(\mathbf{k},z),\hspace{0.3in}\alpha
,\beta =1,2,3.  \label{6.11}
\end{equation}
The integrals $I_{\beta }(\mathbf{k},z)$and $Q_{\alpha \beta }(\mathbf{k},z)$
can be calculated exactly just as for the transverse perturbation above. As
in that case, the poles of $(\mathbf{1}-Q(\mathbf{k},z))^{-1}$gives the
hydrodynamic excitations. In place of (\ref{6.8}), the longitudinal
hydrodynamic modes are now defined by the three solutions to
\begin{equation}
\det (\mathbf{1}-Q(\mathbf{k},z\left( k\right) ))=0.  \label{6.12}
\end{equation}
At $k=0$, the matrix $\mathbf{1}-Q(\mathbf{k},z)$ is found to be
\begin{equation}
\mathbf{1}-Q(\mathbf{0},z)=\left(
\begin{array}{ccc}
\frac{z}{z+\nu } & 0 & 0 \\
-\frac{\nu -2\zeta }{z+\nu -\zeta }+\frac{\nu }{z+\nu } & \frac{z+\frac{
\zeta }{2}}{z+\nu -\zeta } & 0 \\
0 & 0 & \frac{z-\frac{\zeta }{2}}{z+\nu -\frac{\zeta }{2}}
\end{array}
\right) .  \label{6.13}
\end{equation}
The solutions to (\ref{6.12}) give the longitudinal modes in the long
wavelength limit
\begin{equation}
z(0)\rightarrow \left\{ 0,-\frac{\zeta }{2},\frac{\zeta }{2}\right\} .
\label{6.14}
\end{equation}
Comparison with (\ref{4.11}) shows that these are indeed the
hydrodynamic modes.

It is straight forward to extend the calculation of these modes by
expanding $Q(\mathbf{k},z)$ to order $k^{2}$. The solutions to
(\ref{6.12}) are found to be the same as in (\ref{4.10})
\begin{equation}
z(\mathbf{k})\rightarrow \left\{ 0+\frac{1}{\zeta }k^{2},-\frac{\zeta }{2}
+\left( \frac{2}{3\zeta }-\frac{5}{4}\kappa \right) k^{2},\frac{\zeta }{2}
-\left( \frac{1}{3\zeta }+\frac{2}{3}\eta \right) k^{2}\right\} .
\label{6.15}
\end{equation}
The shear viscosity $\eta $ is again given by (\ref{6.10}), while
the thermal conductivity $\kappa $ is
\begin{equation}
\kappa (\alpha )=\frac{2((\nu (\alpha )-\zeta (\alpha ))^{2}+\zeta
^{2}(\alpha ))}{3(\nu (\alpha )-2\zeta (\alpha ))^{2}\nu (\alpha )}.
\label{6.16}
\end{equation}
This agrees in detail with the expression obtained by the Chapman-Enskog
procedure \cite{BMD96}.

The results of this section complete the first objective of this work:
demonstration that the hydrodynamic excitations exist in the solution to the
linear kinetic equation for small perturbations of the HCS. The form of the
hydrodynamic modes to Navier-Stokes order and dependence of the transport
coefficients on the restitution coefficient is confirmed in detail,
providing independent support for the assumptions of the Chapman-Enskog
method. In addition, Figure 1 shows that the hydrodynamic modes exist far
beyond the validity domain for the Navier-Stokes approximation. For the
example considered here that domain appears to be quite restricted.

\section{Approach to hydrodynamics}

The second objective of this work is to verify that the hydrodynamic
excitations are the slowest modes at long wavelengths and therefore dominate
at long times. First, the general solution to the linearized kinetic
equation (\ref{5.11}) is reconsidered to display all of the possible
excitations.
\begin{equation}
\delta \tilde{\tilde{f}}(\mathbf{k},\mathbf{v},z)=\mathcal{R}\delta \tilde{f}
(\mathbf{k},\mathbf{v},0)+\mathcal{R}M_{\alpha }(\mathbf{v})(\mathbf{1}-Q(
\mathbf{k},z))_{\alpha \beta }^{-1}I_{\beta }(\mathbf{k},z).  \label{7.1}
\end{equation}
The dynamics arises from two types of contributions. The first is the
non-hydrodynamic behavior generated by the resolvent operator in $\mathcal{R}
\delta \tilde{f}$, $\mathcal{R}M_{\alpha }$, and $I_{\beta }(\mathbf{k},z)$
(see Eq. (\ref{5.12})). The hydrodynamic exciations arise from the simple
poles of $(\mathbf{1}-Q(\mathbf{k},z))_{\alpha \beta }^{-1}$. Also, there is
a convolution of the hydrodynamic and non-hydrodynamic behavior. All of
these effects are present even for $k\rightarrow 0$, where these expressions
can be made more explicit. Hence the following analysis is first carried out
at $k=0$ and then it is argued that the conclusions extend to $k\neq 0$ by
continuity.

Using (\ref{5.12}) and the results of the previous section at $k=0$ the
general solution can be written exactly
\begin{eqnarray}
\delta \tilde{f}(\mathbf{0},\mathbf{v},s) &=&\sum_{\alpha }e^{z_{\alpha
}(0)s}\chi _{\alpha }(\mathbf{v})\tilde{y}_{\alpha }^{\prime }(\mathbf{0}
,0)+e^{-\left( \frac{3}{2}\zeta +\nu \right) s}\left( \delta \tilde{f}(
\mathbf{0},e^{-\frac{1}{2}\zeta s}\mathbf{v},0)\right.   \nonumber \\
&&\left. +\sum_{\alpha }e^{z_{\alpha }(0)s}\chi _{\alpha }(e^{-\frac{1}{2}
\zeta s}\mathbf{v})y_{\alpha }^{\prime }(0)\right) ,  \label{7.2}
\end{eqnarray}
The new hydrodynamic fields $\left\{ y_{\alpha }^{\prime }\right\} $in (\ref
{7.2}) are linear combinations of $\left\{ y_{\alpha }\right\} $and the
functions $\chi _{\alpha }\left( \mathbf{v}\right) $are the corresponding
linear combinations of $M_{\alpha }(\mathbf{v})$
\begin{equation}
y_{\alpha }^{\prime }=\left\{ \delta n,\delta n+\frac{\delta T}{2},\delta
\mathbf{u}\right\} ,\hspace{0.1in}\chi _{\alpha }=\left\{ \left(
f_{hcs}+\nabla _{\mathbf{v}}\cdot \mathbf{v}f_{hcs}\right) ,-\nabla _{
\mathbf{v}}\cdot \mathbf{v}f_{hcs},\nabla _{\mathbf{v}}f_{hcs}\right\} .
\label{7.3}
\end{equation}
The first term on the right side of (\ref{7.2})is the contribution from hydrodynamic
excitations. It could have been constructed directly from the solution to
the linearized kinetic equation as an eigenvalue problem. Then $\chi
_{\alpha }(\mathbf{v})$ would be the hydrodynamic eigenfunctions. This has
been done in \cite{BDmodes03} and the above $\chi _{\alpha }$ are
indeed the eigenfunctions found there.

All other excitations in the system are contained in the second term of (\ref
{7.2}). Nominally, this appears to decay as $\exp (-\left( \frac{3}{2}\zeta
+\nu \right) s)$. This is misleading, however, due to the additional time
dependence in the functions of velocity. Consider first the moments giving
the hydrodynamic fields $y_{\alpha }^{\prime }(\mathbf{k},s)$
\begin{equation}
y_{\alpha }^{\prime }(\mathbf{k},s)=\int d\mathbf{v}\psi _{\alpha }(\mathbf{
v })\delta \tilde{f}(\mathbf{k},\mathbf{v},s),\hspace{0.3in}\psi _{\alpha
}=\left\{ 1,\frac{1}{3}v^{2}+\frac{1}{2},\mathbf{v}\right\} .  \label{7.4}
\end{equation}
Then (\ref{7.2}) gives
\begin{eqnarray}
y_{\alpha }^{\prime }(\mathbf{0},s) &=&e^{z_{\alpha }(0)s}\tilde{y}_{\alpha
}^{\prime }(\mathbf{0},0)+e^{-\nu s}\int d\mathbf{v}\psi _{\alpha }\left( e^{
\frac{1}{2}\zeta s}\mathbf{v}\right) \left( \delta \tilde{f}(\mathbf{0},
\mathbf{v},0)\right.  \nonumber \\
&&\left. +\sum_{\alpha }e^{z_{\beta }(0)s}\chi _{\beta }(\mathbf{v})y_{\beta
}^{\prime }(\mathbf{0},0)\right)  \label{7.5}
\end{eqnarray}
Since the functions $\psi _{\alpha }\left( \mathbf{v}\right)$ are
polynomials of maximum degree $2$, the second term decays as $\exp (-\left(
\nu -\zeta \right) s)$. The fastest decaying hydrodynamic mode is that with $
z_{2}(0)=-\zeta /2$. Thus the hydrodynamic modes are bounded away from all
of the microscopic modes and the fields $y_{\alpha }^{\prime }( \mathbf{0}
,s)$ obey the hydrodynamic equations for long times, if $\nu >3\zeta /2$.
This must be the case since this condition characterizes energy loss,
according to the moment conditions on the collision integral (\ref{2.2}).
For the model considered, $3\zeta /2=$$5\left( 1-\alpha ^{2}\right) /8$ and $
\nu =1-(1-\alpha )^{2}/4$, so the condition holds even at $\alpha =0$.

Now consider other properties obtained as averages over the distribution
function. In general, for a property determined from $A(\mathbf{v})$

\begin{eqnarray}
\left\langle A;s\right\rangle  &=&\left\langle A(\mathbf{v})\right\rangle
_{hcs}+\sum_{\alpha }e^{z_{\alpha }(0)s}y_{\alpha }^{\prime }(0)\int d%
\mathbf{v}A(\mathbf{v})\chi _{\alpha }(\mathbf{v})  \nonumber \\
&&+e^{-\nu s}\int d\mathbf{v}A(e^{\frac{1}{2}\zeta s}\mathbf{v})\left(
\delta \tilde{f}(\mathbf{0},\mathbf{v},0)+\sum_{\alpha }e^{z_{\alpha
}(0)s}\chi _{\alpha }(\mathbf{v})y_{\alpha }^{\prime }(0)\right) 
\end{eqnarray}%
For any bounded function $\left| A(\mathbf{v})\right| <G$ the last term
decays as $e^{-\nu s}$, and since $\nu >\zeta /2$ the hydrodynamic
contribution dominates for long enough times.  However, a peculiarity occurs
for unbounded functions of sufficiently high degree. Suppose $A(\mathbf{v}%
)\sim v^{p}$ for large $\mathbf{v}$. Then the last integral will have a
contribution behaving as $\exp (-\left( \nu -\frac{p}{2}\zeta \right) s)$.
The maximum value of $p$ is restricted by the condition that the average in
the reference state, $\left\langle A(\mathbf{v})\right\rangle _{hcs}$,
should be finite. This condition gives $p<2\nu /\zeta $, so the exponent for
the second term remains negative. Nevertheless, the decay rate can be made
as small as desired within this restriction. In particular, it can be chosen
to give a decay that is slower than the fastest hydrodynamic modes at $\zeta
/2$. For such properties, the hydrodynamic excitations contribute but they
do not dominate at long times. 

The conclusion of this last paragraph suggests new limitations on
hydrodynamics for granular gases. There are two ways to avoid such a
conclusion. The first is to assume that such properties that behave as $A(%
\mathbf{v})\sim v^{p}$ with $\left( \nu -\frac{p}{2}\zeta \right) <\zeta /2$
are unphysical or uninteresting. For example, at $\alpha =0.9$ this gives $%
p>11$. A more precise avoidance of the problem is to formulate the linear
kinetic theory problem in an appropriate Hilbert space whose scalar product
is defined by%
\[
\left( a,b\right) =\int d\mathbf{v}f_{hcs}\left( \mathbf{v}\right) a^{\ast
}\left( \mathbf{v}\right) b\left( \mathbf{v}\right) .
\]%
The solutions of interest are then $\delta f=f_{hcs}\mathbf{\Delta }$, where 
$\Delta $ must lie in the Hilbert space. Similarly, the properties whose
averages are considered must also be in that Hilbert space. Then square
integrability changes the above restriction on $p$ to $2p<2\nu /\zeta $.
This restores the dominance of hydrodynamics at long times for all
admissible physical properties. The choice of a Hilbert space may seem
artificial since normalization of the distribution function requires only
existence in $L_{1}$. However, for normal gases the Hilbert space has been
justified on the basis of the existence of the H functional, or entropy,
when linearized about the equilibrium state. Perhaps there is a
corresponding functional for granular gases leading to the same conclusion.

\section{Summary and conclusions}

Granular gases are significantly different from normal gases. Energy is not
conserved. So, the homogeneous states for an isolated system are time
dependent and the velocity distribution is not Maxwellian. Such states are
unstable with respect to small spatial perturbations. These peculiarities
have often been cited as reasons to question the relevance of hydrodynamics
for granular gases, or at least to expect severe limitations on the
applicability of such a description. The model kinetic equation considered
here supports all of these peculiarities and therefore provides a good
testing ground for such concerns. Previously, the model has been used to
derive hydrodynamic equations by the Chapman-Enskog method for states with
small spatial gradients. The analysis here is complementary, in the sense
that it leads to the same results under those restricted conditions but
describes as well the case of large gradients and the time scale leading up
to the hydrodynamic stage.

The problem posed here is to solve exactly the linear kinetic equation for
small spatial perturbations of the homogeneous reference state (HCS). This
has been accomplished by first solving the equation for the HCS and then
constructing a solution to the linear equation valid for arbitrary space and
time scales.  It has been shown that hydrodynamic excitations exist in this
general solution by identifying them explicitly at long wavelengths and
assuming an analytic connection to shorter wavelengths. In this way the
Navier-Stokes hydrodynamic excitations are demonstrated explicitly. The
existence of hydrodynamic excitations at shorter wavelengths was illustrated
for the special case of the shear mode.  In the last section above, it has
been shown that these excitations dominate the other dynamics for the
hydrodynamic fields and other average properties after a few collisions. In
summary, it appears from the results of this model kinetic equation that the
existence and conditions for a hydrodynamic description for granular gases
is similar to that for normal gases. This conclusion is restricted to the
case of small perturbations of an isolated gas, and further study is
required for the more general and more interesting cases of boundary driven
systems and external forces.
\section{Acknowledgments}

This research was supported in part by a grant from the U. S. Department of
Energy, DE-FG02ER54677.

\appendix

\section{Solution of the kinetic equation}

The formal solution to the model kinetic equation (\ref{3.6}) is
\begin{eqnarray}
f^{\ast }(v^{\ast },s) &=&e^{-\left( \frac{1}{2}\zeta ^{\ast }\left( 3+
\mathbf{v}^{\ast }\cdot \nabla _{\mathbf{v}^{\ast }}\right) +\nu ^{\ast
}\right) s}f^{\ast }(v^{\ast },0)  \nonumber \\
&&+\int_{0}^{s}ds^{\prime }e^{-\left( \frac{1}{2}\zeta ^{\ast }\left( 3+
\mathbf{v}^{\ast }\cdot \nabla _{\mathbf{v}^{\ast }}\right) +\nu ^{\ast
}\right) \left( s-s^{\prime }\right) }\nu ^{\ast }\left( v^{\ast }\right)
g^{\ast }\left( v^{\ast },s^{\prime }\right)  \label{a1}
\end{eqnarray}
The action of the exponentials in (\ref{a1}) can be determined as follows.
Let us begin by considering the more general case that includes spatial
inhomogeneities. Define a function $X(v^{\ast },k,s)$by
\begin{equation}
X(v^{\ast },k,s)=e^{-\left( \frac{1}{2}\zeta ^{\ast }\left( 3+\mathbf{v}
^{\ast }\cdot \nabla _{\mathbf{v}^{\ast }}\right) +\nu ^{\ast }+i\mathbf{k}
\cdot \mathbf{v}\right) s}X\left( v^{\ast }\right)  \label{a2}
\end{equation}
which then obeys the equation
\begin{equation}
\left( \partial _{s}+\frac{1}{2}\zeta ^{\ast }\left( 3+\mathbf{v}^{\ast
}\cdot \nabla _{\mathbf{v}^{\ast }}\right) +\nu ^{\ast }+i\mathbf{k}\cdot
\mathbf{v}^{\ast }\right) X=0  \label{a3}
\end{equation}
Next introduce
\begin{equation}
X(v^{\ast },k,s)=e^{-\frac{1}{2}\zeta ^{\ast }s\mathbf{v}^{\ast }\cdot
\nabla _{\mathbf{v}^{\ast }}}\overline{X}(v^{\ast },k,s)  \label{a4}
\end{equation}
so that $\overline{X}(v^{\ast },s)$obeys the equation\
\begin{equation}
\left( \partial _{s}+\frac{3}{2}\zeta ^{\ast }+e^{\frac{1}{2}\zeta ^{\ast }s
\mathbf{v}^{\ast }\cdot \nabla _{\mathbf{v}^{\ast }}}\left( \nu ^{\ast }+i
\mathbf{k}\cdot \mathbf{v}^{\ast }\right) e^{-\frac{1}{2}\zeta ^{\ast }s
\mathbf{v}^{\ast }\cdot \nabla _{\mathbf{v}^{\ast }}}\right) \overline{X}=0.
\label{a5}
\end{equation}
From the identity
\begin{equation}
e^{\frac{1}{2}\zeta ^{\ast }s\mathbf{v}^{\ast }\cdot \nabla _{\mathbf{v}
^{\ast }}}F(v^{\ast })e^{-\frac{1}{2}\zeta ^{\ast }s\mathbf{v}^{\ast }\cdot
\nabla _{\mathbf{v}^{\ast }}}=F\left( e^{\frac{1}{2}\zeta ^{\ast }s}v^{\ast
}\right)  \label{a6}
\end{equation}
this equation becomes
\begin{equation}
\left( \partial _{s}+\frac{3}{2}\zeta ^{\ast }+\nu ^{\ast }+e^{\frac{1}{2}
\zeta ^{\ast }s}i\mathbf{k}\cdot \mathbf{v}^{\ast }\right) \overline{X}=0
\label{a7}
\end{equation}
This can be integrated directly and inserted in (\ref{a4}) to give
\[
X\left( v^{\ast },s\right) =e^{-\frac{3}{2}\zeta ^{\ast }s}e^{-\frac{1}{2}%
\zeta ^{\ast }s\mathbf{v}^{\ast }\cdot \nabla _{\mathbf{v}^{\ast
}}}e^{-\nu ^{\ast }s}\exp \left( -\int_{0}^{s}ds^{\prime
}e^{\frac{1}{2}\zeta ^{\ast }s^{\prime }}i\mathbf{k}\cdot
\mathbf{v}^{\ast }\right) X\left( v^{\ast }\right)
\]%
\begin{equation}
=e^{-\frac{3}{2}\zeta ^{\ast }s}e^{-\nu ^{\ast }s}\exp \left(
-\int_{0}^{s}ds^{\prime }e^{-\frac{1}{2}\zeta ^{\ast }s^{\prime }}i\mathbf{k}%
\cdot \mathbf{v}^{\ast }\right) X\left( e^{-\frac{1}{2}\zeta
^{\ast }s}v^{\ast }\right)   \label{a8}
\end{equation}
In particular, if $k=0,$the formal solution to the kinetic equation for a
homogeneous system becomes
\begin{eqnarray}
f^{\ast }(\mathbf{v}^{\ast },s) &=&e^{-\frac{3}{2}\zeta ^{\ast }s}e^{-\nu
^{\ast }s}f^{\ast }(e^{-\frac{1}{2}\zeta ^{\ast }s}\mathbf{v}^{\ast },0)
\nonumber \\
&&+\int_{0}^{s}ds^{\prime }e^{-\frac{3}{2}\zeta ^{\ast }s^{\prime }}e^{-\nu
^{\ast }s^{\prime }}\nu ^{\ast }g^{\ast }(e^{-\frac{1}{2}\zeta ^{\ast
}s^{\prime }}\mathbf{v}^{\ast },s-s^{\prime }),  \label{a9}
\end{eqnarray}

\section{Linearization of the model kinetic operator}

The collision operator Eq.(\ref{2.10}) to linear order in the
small deviation $\delta f$ becomes
\begin{equation}
J[f]\rightarrow -\nu _{hcs}\left( f_{hcs}-g_{hcs}\right) -\nu
_{hcs}\left( \delta f-\delta g\right) -\delta \nu \left(
f_{hcs}-g_{hcs}\right) \label{a.0}
\end{equation}%
Now, since $\nu $ and $g$ are normal
\begin{eqnarray}
\delta g &=&\delta y_{\alpha }\frac{\partial g}{\partial y_{\alpha
}}\mid
_{hcs}=\delta y_{\alpha }\frac{\partial g_{hcs}}{\partial y_{\alpha }}\mid _{%
\mathbf{V}=\mathbf{v}},  \nonumber \\
\hspace{0.3in}\delta \nu  &=&\delta y_{\alpha }\frac{\partial \nu
}{\partial
y_{\alpha }}\mid _{hcs}=\delta y_{\alpha }\frac{\partial \nu _{hcs}}{%
\partial y_{\alpha }}\mid _{\mathbf{V}=\mathbf{v}}  \label{a.01}
\end{eqnarray}%
where $y_{\alpha }$'s are the hydrodynamic fields as before. Note
that the
derivatives must be taken using the $HCS$ as a function of $\mathbf{V}=%
\mathbf{v-u}$, and then setting $\mathbf{u}=0$. The kinetic
equation becomes
\begin{eqnarray}
\left( \partial _{t}+\mathbf{v\cdot \nabla +}\nu _{hcs}\right)
\delta f &=&-\delta y_{\alpha }\left\{ \frac{\partial }{\partial
y_{\alpha }}\left[ \nu \left( f-g\right) \right] \right\} \mid
_{hcs}+\nu _{hcs}\delta
y_{\alpha }\frac{\partial f_{hcs}}{\partial y_{\alpha }}  \nonumber \\
&=&\delta y_{\alpha }\left[ \frac{\partial }{\partial y_{\alpha }}\frac{1}{2}%
\zeta _{hcs}\nabla _{\mathbf{V}}\cdot \left(
\mathbf{V}f_{hcs}\right) \right]
_{\mathbf{V}=\mathbf{v}}  \nonumber \\
&&+\nu _{hcs}\delta y_{\alpha }\frac{\partial f_{hcs}}{\partial
y_{\alpha }}. \label{a.1}
\end{eqnarray}%
Use has been made of the fact that the HCS obeys the equation
\begin{equation}
\frac{1}{2}\zeta _{hcs}\nabla _{\mathbf{V}}\cdot \left( \mathbf{V}%
f_{hcs}\right) =\nu \left( f_{hcs}-g_{hcs}\right) .  \label{a.2}
\end{equation}%
in the second equality. More explicitly, carrying out the
derivatives using the variables $n,T,$and $u_{i}$
\begin{eqnarray}
\left( \partial _{t}+\mathbf{v\cdot \nabla +}\nu _{hcs}\right)
\delta f &=&\delta n\frac{\partial }{\partial n}\left[
\frac{1}{2}\zeta _{hcs}\nabla
_{\mathbf{V}}\cdot \left( \mathbf{V}f_{hcs}\right) \right] _{\mathbf{V}=%
\mathbf{v}}  \nonumber \\
&&+\delta \mathbf{u}\cdot \nabla _{\mathbf{u}}\left[
\frac{1}{2}\zeta
^{(0)}\nabla _{\mathbf{V}}\cdot \left( \mathbf{V}f_{hcs}\right) \right] _{%
\mathbf{V}=\mathbf{v}}  \nonumber \\
&&+\delta T\frac{\partial }{\partial T}\left[ \frac{1}{2}\zeta
^{(0)}\nabla
_{\mathbf{V}}\cdot \left( \mathbf{V}f_{hcs}\right) \right] _{\mathbf{V}=%
\mathbf{v}}  \nonumber \\
&&+\nu _{hcs}\left( \delta n\frac{\partial f_{hcs}}{\partial n}+\delta u_{i}%
\frac{\partial f_{hcs}}{\partial u_{i}}+\delta T\frac{\partial f_{hcs}}{%
\partial T}\right)   \label{a.3}
\end{eqnarray}%
Using the normal form of the local HCS these derivatives can be
evaluated to give
\begin{eqnarray}
\left( \partial _{t}+\mathbf{v\cdot \nabla +}\nu _{hcs}\right)
\delta f &=&\left( \zeta _{hcs}\nabla _{\mathbf{v}}\cdot
\mathbf{v}f_{hcs}\right)
\frac{\delta n}{n_{hcs}}  \nonumber \\
&&+\frac{1}{4}\zeta _{hcs}\left( \nabla _{\mathbf{v}}\cdot \mathbf{v}%
f_{hcs}-\nabla _{\mathbf{v}}\cdot \mathbf{v}\nabla
_{\mathbf{v}}\cdot
\mathbf{v}f_{hcs}\right) \frac{\delta T}{T_{hcs}}  \nonumber \\
&&-\frac{1}{2}\zeta _{hcs}\left( \nabla _{\mathbf{v}}f_{hcs}+\nabla _{%
\mathbf{v}}\cdot \mathbf{v}\nabla _{\mathbf{v}}f_{hcs}\right)
\cdot \delta
\mathbf{u}+\nu _{hcs}f_{hcs}\frac{\delta n}{n_{hcs}}  \nonumber \\
&&-\frac{\nu _{hcs}}{2}\nabla _{\mathbf{v}}\cdot \mathbf{v}f_{hcs}\frac{%
\delta T}{T_{hcs}}-\nu _{hcs}\nabla _{\mathbf{v}}f_{hcs}\cdot \delta \mathbf{%
u}  \label{a.4}
\end{eqnarray}%
This equation in scaled variables becomes
\begin{equation}
\left( \partial _{s}+\frac{1}{2}\zeta ^{\ast }\nabla
_{\mathbf{v}^{\ast }}\cdot \mathbf{v}^{\ast }+v^{\ast }\cdot
\nabla _{\mathbf{r}^{\ast }}+\nu ^{\ast }\right) \delta f^{\ast
}=M_{\alpha }^{\ast }(\mathbf{v}^{\ast })y_{\alpha }^{\ast
}(\mathbf{r}^{\ast },s)  \label{a.5}
\end{equation}%
where $M_{\alpha }^{\ast }(\mathbf{v}^{\ast })$ and $y_{\alpha }^{\ast }(%
\mathbf{r}^{\ast },s)$ are as defined in Eq.(\ref{5.3}) and Eq.(
\ref{4.4}), respectively. This gives the linearized dynamics of
the system for small deviations from the HCS.

\section{Evaluation of $I$ and $Q$}

Here, the action of the resolvent operator defined in Eq.
(\ref{5.6}) of the text is made explicit. In particular, the
functions
\begin{equation}
I_{\alpha }(\mathbf{k},z)=\int d\mathbf{v}\phi _{\alpha }(\mathbf{v})%
\mathcal{R}\delta
\tilde{f}(\mathbf{k},\mathbf{v},0),\hspace{0.4cm}Q_{\alpha
\beta }(\mathbf{k},z)=\int d\mathbf{v}\phi _{\alpha }(\mathbf{v})\mathcal{R}%
M_{\beta }.  \label{C1}
\end{equation}%
are evaluated. Note that scaled variables are used and the
superscript asterisks have been suppressed. First, define a
Fourier transform with respect to the velocity $\mathbf{v}$ of the
form
\begin{equation}
\widehat{Y}(\mathbf{\lambda })=\int d\mathbf{v}e^{i\mathbf{\lambda \cdot v}%
}Y(\mathbf{v}).  \label{C2}
\end{equation}%
Then, Eq.(\ref{C1}) above can be written as
\begin{equation}
I_{\alpha }(\mathbf{k},z)=\left( 2\pi \right) ^{-3}\int d\mathbf{\lambda }%
\delta \widehat{\tilde{f}}_{t}(\mathbf{k},\mathbf{\lambda },0)\int d\mathbf{v%
}\phi _{\alpha }\mathcal{R}e^{-i\mathbf{\lambda \cdot v}},
\label{C3}
\end{equation}%
and
\begin{equation}
Q_{\alpha \beta }(\mathbf{k},z)=\left( 2\pi \right) ^{-3}\int d\mathbf{%
\lambda }\widehat{M}_{\beta }\left( \mathbf{\lambda }\right) \int d\mathbf{v}%
\phi _{\alpha }\mathcal{R}e^{-i\mathbf{\lambda \cdot v}}
\label{C4}
\end{equation}%
Hence in general it is sufficient to describe the action of the
resolvant on the Fourier mode function $e^{-i\mathbf{\lambda \cdot
v}}$.

It follows from (\ref{a2}) and (\ref{a8}) Appendix A above that
\begin{eqnarray}
\mathcal{R}e^{-i\mathbf{\lambda \cdot v}} &=&\int_{0}^{\infty
}dse^{-\left(
z+\nu +i\mathbf{k}\cdot \mathbf{v}+\frac{1}{2}\zeta \nabla _{\mathbf{v}%
}\cdot \mathbf{v}\right) s}e^{-i\mathbf{\lambda \cdot v}}  \nonumber \\
&=&\int_{0}^{\infty }dse^{-zs}e^{-\frac{3}{2}\zeta s}e^{-vs}\exp \left( -i%
\mathbf{k(}s)\cdot \mathbf{v}\right) e^{-ie^{-\frac{1}{2}\zeta s}\mathbf{%
\lambda \cdot v}},  \label{C5}
\end{eqnarray}%
with
\begin{equation}
\mathbf{k(}s)=\int_{0}^{s}ds^{\prime }e^{-\frac{1}{2}\zeta s^{\prime }}%
\mathbf{k.}  \label{C6}
\end{equation}%
It follows that
\begin{eqnarray}
\int d\mathbf{v}\phi _{\alpha }\mathcal{R}e^{-i\mathbf{\lambda
\cdot v}}
&=&\int_{0}^{\infty }dse^{-\left( z+v\right) s}e^{-\frac{3}{2}\zeta s}\int d%
\mathbf{v}\phi _{\alpha }\left( \mathbf{v}\right)
e^{-i\mathbf{\lambda \cdot }e^{-\frac{1}{2}\zeta
s}\mathbf{v}}e^{-i\mathbf{k(}s)\cdot \mathbf{v}}
\nonumber \\
&=&\int_{0}^{\infty }dse^{-\left( z+v\right) s}\int
d\mathbf{v}\phi _{\alpha
}\left( e^{\frac{1}{2}\zeta s}\mathbf{v}\right) e^{i\left( \mathbf{k(-}s)-%
\mathbf{\lambda }\right) \cdot \mathbf{v}}  \label{C7}
\end{eqnarray}%
Putting in the forms of the $\left\{ \phi _{\alpha }\right\} $  given in Eq.(%
\ref{5.8} ) leads to
\begin{equation}
\int d\mathbf{v}\phi _{1}\mathcal{R}e^{-i\mathbf{\lambda \cdot v}%
}=\int_{0}^{\infty }dse^{-\left( z+v\right) s}\left( 2\pi \right)
^{3}\delta \left( \mathbf{k(-}s)-\mathbf{\lambda }\right)
\label{C9}
\end{equation}%
\begin{equation}
\int d\mathbf{v}\phi _{2}\mathcal{R}e^{-i\mathbf{\lambda \cdot v}%
}=\int_{0}^{\infty }dse^{-\left( z+v\right) s}\left(
-\frac{2}{3}e^{\zeta s}\nabla _{\lambda }^{2}-1\right) \left( 2\pi
\right) ^{3}\delta \left( \mathbf{k(-}s)-\mathbf{\lambda }\right)
\label{C10}
\end{equation}%
\begin{equation}
\int d\mathbf{v}\phi _{3}\mathcal{R}e^{-i\mathbf{\lambda \cdot v}}=i\widehat{%
\mathbf{k}}\cdot \nabla _{\lambda }\int_{0}^{\infty }dse^{-\left( z+v-\frac{1%
}{2}\zeta \right) s}\left( 2\pi \right) ^{3}\delta \left( \mathbf{k(-}s)-%
\mathbf{\lambda }\right)   \label{C11}
\end{equation}%
\begin{equation}
\int d\mathbf{v}\phi _{4,5}\mathcal{R}e^{-i\mathbf{\lambda \cdot v}}=i%
\widehat{\mathbf{e}}_{1,2}\cdot \nabla _{\lambda }\int_{0}^{\infty
}dse^{-\left( z+v-\frac{1}{2}\zeta \right) s}\left( 2\pi \right)
^{3}\delta \left( \mathbf{k(-}s)-\mathbf{\lambda }\right)
\label{C11a}
\end{equation}%
\qquad \qquad

This allows evaluation of $I_{\alpha }(\mathbf{k},z)$
\begin{equation}
I_{1}(\mathbf{k},z)=\int_{0}^{\infty }dse^{-\left( z+v^{\ast
}\right) s}\delta
\widehat{\tilde{f}}_{t}(\mathbf{k},\mathbf{k(-}s),0),  \label{C12}
\end{equation}%
\begin{equation}
I_{2}(\mathbf{k},z)=\int_{0}^{\infty }dse^{-\left( z+v^{\ast
}\right)
s}\left( -\frac{2}{3}e^{\zeta ^{\ast }s}\nabla _{\mathbf{k(-}%
s)}^{2}-1\right) \delta
\widehat{\tilde{f}}_{t}(\mathbf{k},\mathbf{k(-}s),0), \label{C13}
\end{equation}%
\begin{equation}
I_{3}(\mathbf{k},z)=\int_{0}^{\infty }dse^{-\left( z+v^{\ast }-\frac{1}{2}%
\zeta ^{\ast }\right) s}i\widehat{\mathbf{k}}\cdot \nabla _{\mathbf{k(-}%
s)}\delta \widehat{\tilde{f}}_{t}(\mathbf{k},\mathbf{k(-}s),0),
\label{C14}
\end{equation}%
\begin{equation}
I_{4,5}(\mathbf{k},z)=\int_{0}^{\infty }dse^{-\left( z+v^{\ast }-\frac{1}{2}%
\zeta ^{\ast }\right) s}i\widehat{\mathbf{e}}_{1,2}\cdot \nabla _{\mathbf{k(-%
}s)}\delta \widehat{\tilde{f}}_{t}(\mathbf{k},\mathbf{k(-}s),0),
\label{C15}
\end{equation}%
The elements $Q_{\alpha \beta }(\mathbf{k},z)$ are evaluated in a similar way%
\begin{equation}
Q_{1\beta }(\mathbf{k},z)=\int_{0}^{\infty }dse^{-\left( z+v\right) s}%
\widehat{M}_{\beta }\left( \mathbf{k(-}s)\right)
\end{equation}%
\begin{equation}
Q_{2\beta }(\mathbf{k},z)=\int_{0}^{\infty }dse^{-\left(
z+v\right) s}\left(
-\frac{2}{3}e^{\zeta s}\nabla _{\mathbf{k}(-s)}^{2}-1\right) \widehat{M}%
_{\beta }\left( \mathbf{k(-}s)\right)
\end{equation}%
\begin{equation}
Q_{3\beta }(\mathbf{k},z)=\int_{0}^{\infty }dse^{-\left( z+v-\frac{1}{2}%
\zeta \right) s}i\widehat{\mathbf{k}}\cdot \nabla _{\mathbf{k(-}s)}\widehat{M%
}_{\beta }\left( \mathbf{k(-}s)\right)
\end{equation}%
\begin{equation}
Q_{4,5\beta }(\mathbf{k},z)=\int_{0}^{\infty }dse^{-\left( z+v-\frac{1}{2}%
\zeta \right) s}i\widehat{\mathbf{e}}_{1,2}\cdot \nabla _{\mathbf{k(-}s)}%
\widehat{M}_{\beta }\left( \mathbf{k(-}s)\right)
\end{equation}%
The $\widehat{M}_{\beta }\left( \mathbf{\lambda }\right) $ are
obtained
directly from the definition of $\left\{ M_{\beta }\right\} $ in Eq.(\ref%
{5.3})
\begin{equation}
\widehat{M}_{\beta }(\mathbf{\lambda })=\left(
\begin{array}{c}
\left( -\zeta \mathbf{\lambda }\cdot \nabla _{\mathbf{\lambda
}}+\nu \right)
\widehat{f}_{hcs}\left( \lambda \right)  \\
-\frac{1}{2}\mathbf{\lambda }\cdot \nabla _{\mathbf{\lambda }}\left( \frac{1%
}{2}\zeta -\left( -\frac{1}{2}\zeta \mathbf{\lambda }\cdot \nabla _{\mathbf{%
\lambda }}+\nu \right) \right) \widehat{f}_{hcs}\left( \lambda \right)  \\
\left( \frac{1}{2}\zeta \left( 1-\mathbf{\frac{1}{2}\lambda }\cdot \nabla _{%
\mathbf{\lambda }}\right) +\nu \right) i\mathbf{\lambda }\widehat{f}%
_{hcs}\left( \lambda \right)
\end{array}%
\right)   \label{C16}
\end{equation}%
where
\begin{eqnarray}
\widehat{f}_{hcs}\left( \lambda \right)  &=&\int d\mathbf{v}e^{i\mathbf{%
\lambda \cdot v}}f_{hcs}\left( v\right)   \label{C17} \\
&=&\int d\mathbf{v}e^{i\mathbf{\lambda \cdot v}}\int_{0}^{\infty
}ds^{\prime
}e^{-\frac{3}{2}\zeta s^{\prime }}e^{-\nu s^{\prime }}\nu \left( \frac{B}{%
\pi }\right) ^{3/2}e^{-B\left( e^{-\frac{1}{2}\zeta s^{\prime
}}v\right)
^{2}}  \nonumber \\
&=&\nu \int_{0}^{\infty }ds^{\prime }e^{-\nu s^{\prime }}\allowbreak e^{-%
\frac{1}{4}\mathbf{\lambda }^{2}\left( s^{\prime }\right) }.
\nonumber
\end{eqnarray}%
with the definition
\begin{equation}
\mathbf{\lambda }(s^{\prime })=\frac{e^{\frac{1}{2}\zeta s^{\prime }}}{\sqrt{%
B}}\mathbf{\lambda .}  \label{C18}
\end{equation}

As an example, $Q_{44}(\mathbf{k},z)$ for the transverse shear
mode is
\begin{equation}
Q_{44}(\mathbf{k},z)=\int_{0}^{\infty }dse^{-\left(
z+v-\frac{1}{2}\zeta \right) s}\left(
-i\widehat{\mathbf{e}}_{1}\cdot \nabla _{\lambda }\left(
\frac{1}{2}\zeta \left( 1-\frac{1}{2}\mathbf{\lambda }\cdot \nabla _{\mathbf{%
\lambda }}\right) -\nu \right) \right.   \label{C19}
\end{equation}%
\[
\left. \times i\widehat{\mathbf{e}}_{1}\cdot \mathbf{\lambda }\widehat{f}%
_{hcs}\left( \lambda \right) \right) \mid _{\mathbf{\lambda
=k(-}s)}
\]%
\begin{equation}
=\left( \nu -\zeta k^{2}\frac{d}{dk^{2}}\right) \int_{0}^{\infty
}dse^{-\left( z+\nu -\frac{1}{2}\zeta \right) s}\nu
\int_{0}^{\infty }ds^{\prime }e^{-\nu s^{\prime }}\allowbreak \exp
\left( -e^{\zeta s^{\prime }}a(s)k^{2}\right)   \label{C20}
\end{equation}%
where
\begin{equation}
a(s)=\left( 1-\frac{\zeta }{\nu }\right) \left(
\frac{e^{\frac{1}{2}\zeta s}-1}{\zeta }\right) ^{2}.  \label{C21}
\end{equation}%
This is the result given in the text.

\bigskip

%BibTeX users please use
%\bibliographystyle{}
%\bibliography{}
%
% Non-BibTeX users please follow the syntax
% the syntax of "referenc.tex" for your own citations
%%%%%%%%%%%%%%%%%%%%%%%% referenc.tex %%%%%%%%%%%%%%%%%%%%%%%%%%%%%%
% sample references
%
%
% Use this file as a template for your own input.
%
%%%%%%%%%%%%%%%%%%%%%%%% Springer-Verlag %%%%%%%%%%%%%%%%%%%%%%%%%%

%
% BibTeX users please use
% \bibliographystyle{}
% \bibliography{}

%%%%%%%%%%%%%%%%%%%%%%%%%%%%%%%%%%%%%%%%%%%%%%%%%%%%%%%%%%%%%%%%%%%%%%

%%%%%%%%%%%%%%%%%%%%%%%%%%%%%%%%%%%%%%%%%%%%%%%%%%%%%%%%%%%%%%%%%%%%%%

\printindex

\end{document}